\documentclass[preprint,12pt]{elsarticle}

\usepackage{amssymb}
\usepackage{amsmath}
\usepackage{xspace}
\usepackage{subfig}
\usepackage{color}
\usepackage{graphicx}
\usepackage{multirow}
\usepackage{multicol}
\usepackage{lineno}
\usepackage{units}
\usepackage{cleveref}
\usepackage{lineno}
\usepackage{soul}

\usepackage{natbib}
\def \gcm{\ifmmode{\mathrm{g/cm}^2}\else{g/cm$^2$}\fi\xspace} 
\def \xmax{\ifmmode{\mathrm{X}_\mathrm{max}} \else{X$_\mathrm{max}$}\fi\xspace} 
\def \xmaxsim{\ifmmode{\mathrm{X}_\mathrm{max}^\mathrm{sim}} \else{X$_\mathrm{max}^\mathrm{sim}$}\fi\xspace} 
\def \xmaxrec{\ifmmode{\mathrm{X}_\mathrm{max}^\mathrm{rec}} \else{X$_\mathrm{max}^\mathrm{rec}$} \fi\xspace} 
\def \deg{\ifmmode{^{\circ}}\else{$^{\circ}$}\fi\xspace} 

\journal{Astroparticle Physics}

\begin{document}

\begin{frontmatter}


\title{Cosmic-ray measurements with an array of Cherenkov telescopes using reconstruction of longitudinal profiles of air showers}

\author[ifsc]{Andr\'es G. Delgado Giler}
\ead{andres.delgado@usp.br}
\author[ifsc]{Vitor de Souza}
\ead{vitor@ifsc.usp.br}

\address[ifsc]{Instituto de F\'isica de S\~ao Carlos, Universidade de S\~ao Paulo, S\~ao Carlos, Brazil.}

\begin{abstract}
  We present a method to reconstruct the longitudinal profile of electrons in showers using Cherenkov telescopes. We show how the Cherenkov light collected by an array of telescopes can be transformed into the number of electrons as a function of atmospheric depth. This method is validated using air shower and simplified telescope simulations. The reconstruction of the depth in which the shower has the maximum number of electrons (\xmax) opens the possibility of cosmic ray composition studies with Cherenkov telescopes in the energy range from 10 to 100 TeV. A resolution of less than 16 \gcm in the \xmax reconstruction is obtained.
\end{abstract}

\begin{keyword}
  Air shower \sep shower maximum \sep Cherenkov telescope.
\end{keyword}

\end{frontmatter}

\section{Introduction}

Cherenkov telescopes are mainly used for gamma-ray astronomy. Non-gamma-ray events have not been fully analysed by standard analysis chains. Exceptions are the study of electron~\cite{bib:electrons:hess,bib:electrons:magic,bib:eletrons:veritas} and extreme (proton and iron) nuclei primaries~\cite{bib:veritas:iron,bib:hess:iron,bib:Jankowsky2020}. Besides those, Cherenkov telescopes measure at least 1000 times more cosmic rays including all nuclei than gamma-ray showers.

This paper proposes a technique to reconstruct the longitudinal profile of electrons in the shower using the signal detected by Cherenkov telescopes. The reconstruction of the longitudinal profile of electrons in the shower could be applied to better understand the shower development in comparison to Monte Carlo simulation and also to improve reconstruction chains. We focus on the possibility to reconstruct the depth in which showers have the maximum number of electrons (\xmax) and its use to determine the primary particle type.

Previous papers, based on the lateral distribution of Cherenkov light produced in the shower, proposed methods to study the cosmic ray composition using imaging atmospheric Cherenkov telescopes (IACTs)~\cite{bib:mass:hegra,KIEDA2001287}. The determination of the \xmax was also studied in references~\cite{bib:PARSONS201426,bib:DENAUROIS2009231}. In these studies, a complex full reconstruction of the shower is impleted via fit of the camera image or deep learning techniques. All parameters are reconstructed together, including \xmax. The complexity of these methods increases significantly with the number of telescopes in the analysis. They also depends on large sets of simulated showers. In reference~\cite{bib:PARSONS201426}, an \xmax resolution of 30 \gcm for energies below 1 TeV was achieved.

In a previous publication~\cite{Giler:2021hvw}, we showed how to use telescopes outside the Cherenkov beam ($> 150$ m) to reconstruct \xmax. In this paper, we make use of the parametrization of the angular distribution function of Cherenkov photons~\cite{Arbeletche:2020rev} to reconstruct the longitudinal profile and calculate \xmax using telescopes at all distances from the shower axis. This is one of the key differences between the method proposed here and previous publications. The new method allows a more direct measurement of the shower properties and the use of telescopes with larger signals and therefore higher probability to trigger. The method proposed here also reduces significantly the complexity of the analysis by focusing in only one parameter (\xmax) which might lead to computational advantages when analising multiple telescopes at the same time.

In section~\ref{sec:simulation}, the shower and the telescope simulation are explained. In section~\ref{sec:reconstruction}, we present the method to reconstruct the longitudinal profile of electrons and the fitting to calculate its maximum. In section~\ref{sec:results}, the reconstructed shower maximum \xmax and the resolution is discussed; and finally in section~\ref{sec:conclusions} the conclusions and final remarks are given.

\section{Simulation of the Cherenkov light reaching the telescopes}
\label{sec:simulation}

We simulated air showers using the CORSIKA 7.6900 package~\cite{1998cmcc.book.....H}. We considered vertical showers generated by gamma, proton and iron nuclei with energy of 10, 30 and 100 TeV. QGSJET II-04~\cite{bib:qgsjet} and URQMD~\cite{bib:urqmd} were used for high and low energy hadronic interactions. The hadronic interaction models do not influence the reconstruction of \xmax which is the focus of this paper. The hadronic interaction models influence the interpretation of \xmax as composition. For each energy and primary particle, 10000 events were simulated. The threshold energies for hadrons, muons, electrons and photons were set to 0.3, 0.01, 0.020, 0.020 GeV, respectively. The U.S. standard atmosphere model was used.

Cherenkov photons were produced in the simulation. Photons were emitted in bunch size 5 with wavelength between 200 and 700 nm. The emission angle of the Cherenkov light is taken to be wavelength-independent and photons are propagated until sea level. An array in the shape of a North-South cross with 24 telescopes delimited by spheres with 2.15 m of radius spaced 40 m apart from each other. All photons reaching the detectors have direction and momentum recorded.

\begin{figure}[]
  \centering
  \includegraphics[scale=0.4]{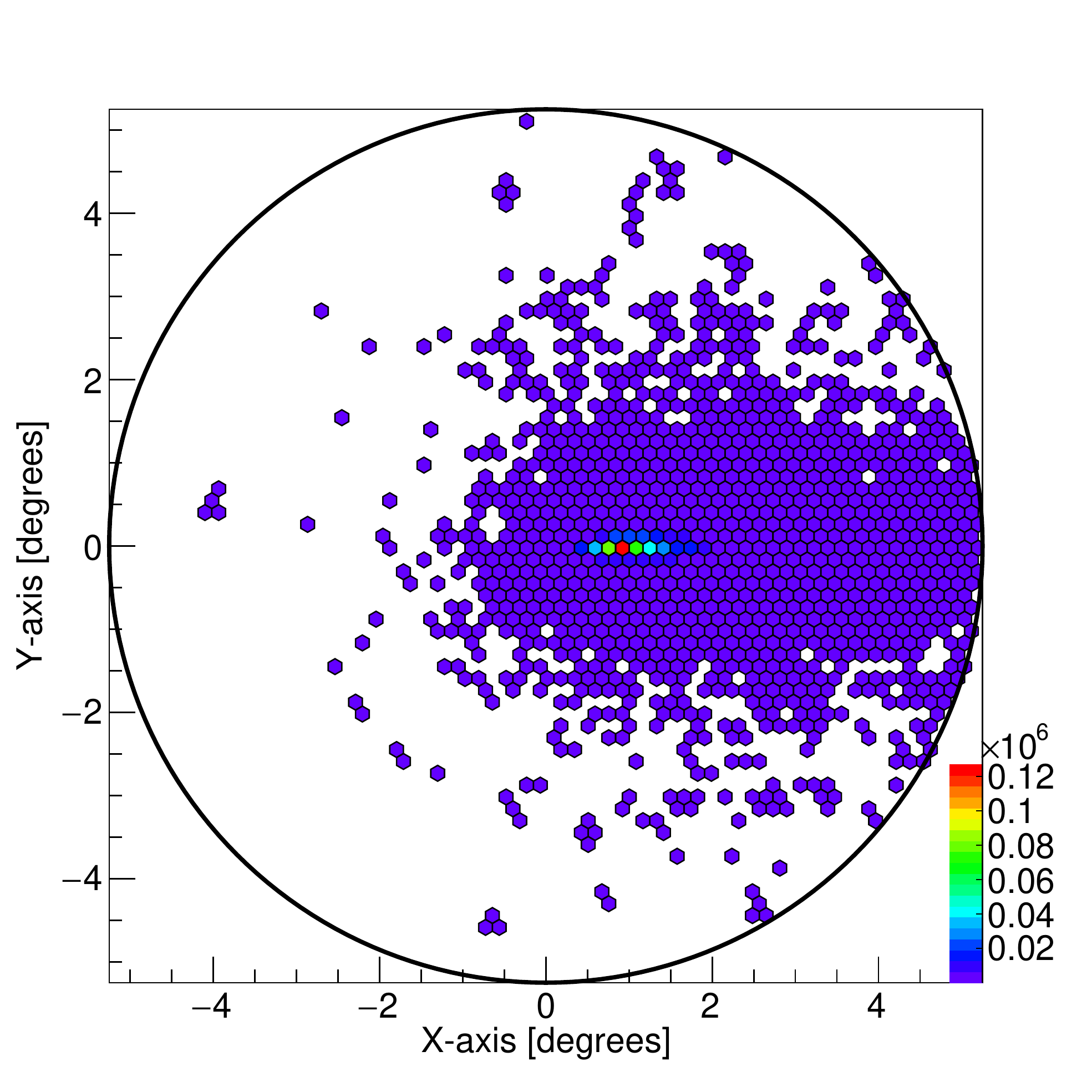}
  \caption{Simplified simulation of a camera image generated by a vertical proton of 100 TeV. The image corresponds to the telescope at 120 m from shower axis. Only pixels satisfying the trigger condition are plotted.}
  \label{fig:camera}
\end{figure}

A simplified simulation of the telescope was considered taking into account the field of view of 10.5 degrees and the camera pixelization with 0.19 degrees hexagonal pixels. Pixels were considered to be triggered if the number of collected photons is three times larger than the night sky background of 2.6$\times$10$^{8}$ photons per sr s m$^{2}$. The sampling rate was taken to be 500 MHz. An example of a simulated image is shown in figure~\ref{fig:camera}.

\section{Reconstruction of the longitudinal profile}
\label{sec:reconstruction}

\begin{figure}[]
  \centering
  \includegraphics[scale=0.35]{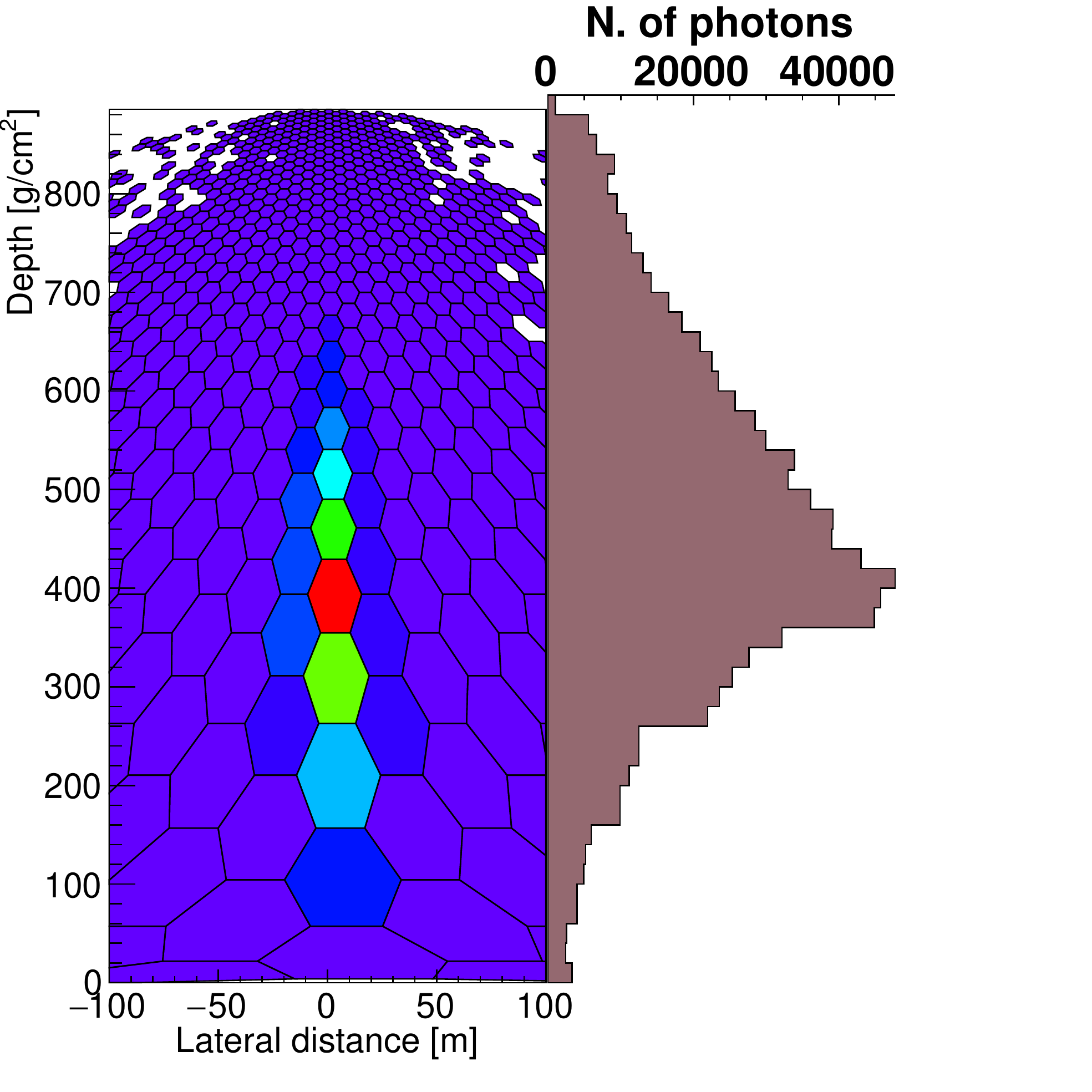}
  \caption{Projected pixels correspond to figure~\ref{fig:camera}. Only pixels that satisfy good projection are plotted. On the right side is shown the number of photons summed along the longitudinal axis.}
  \label{fig:shower:plane}
\end{figure}

The shower plane is defined by the shower axis and the vector pointing to the telescope from the impact point of the shower axis. Each pixel intersects the shower plane defining a region. In a good approximation, all photons detected by one pixel can be considered to be generated in this region. Figure~\ref{fig:shower:plane} shows the number of photons in each region defined by the pixels shown in figure~\ref{fig:camera} and its subsequent (right panel) projection into the shower axis to find the longitudinal profile of Cherenkov photons. A constant density of photons was considered inside each pixel and the photons are distributed into depth bins uniformly. For details about the projection see reference~\cite{Giler:2021hvw}.

\begin{figure}[]
  \centering
  \includegraphics[scale=0.55]{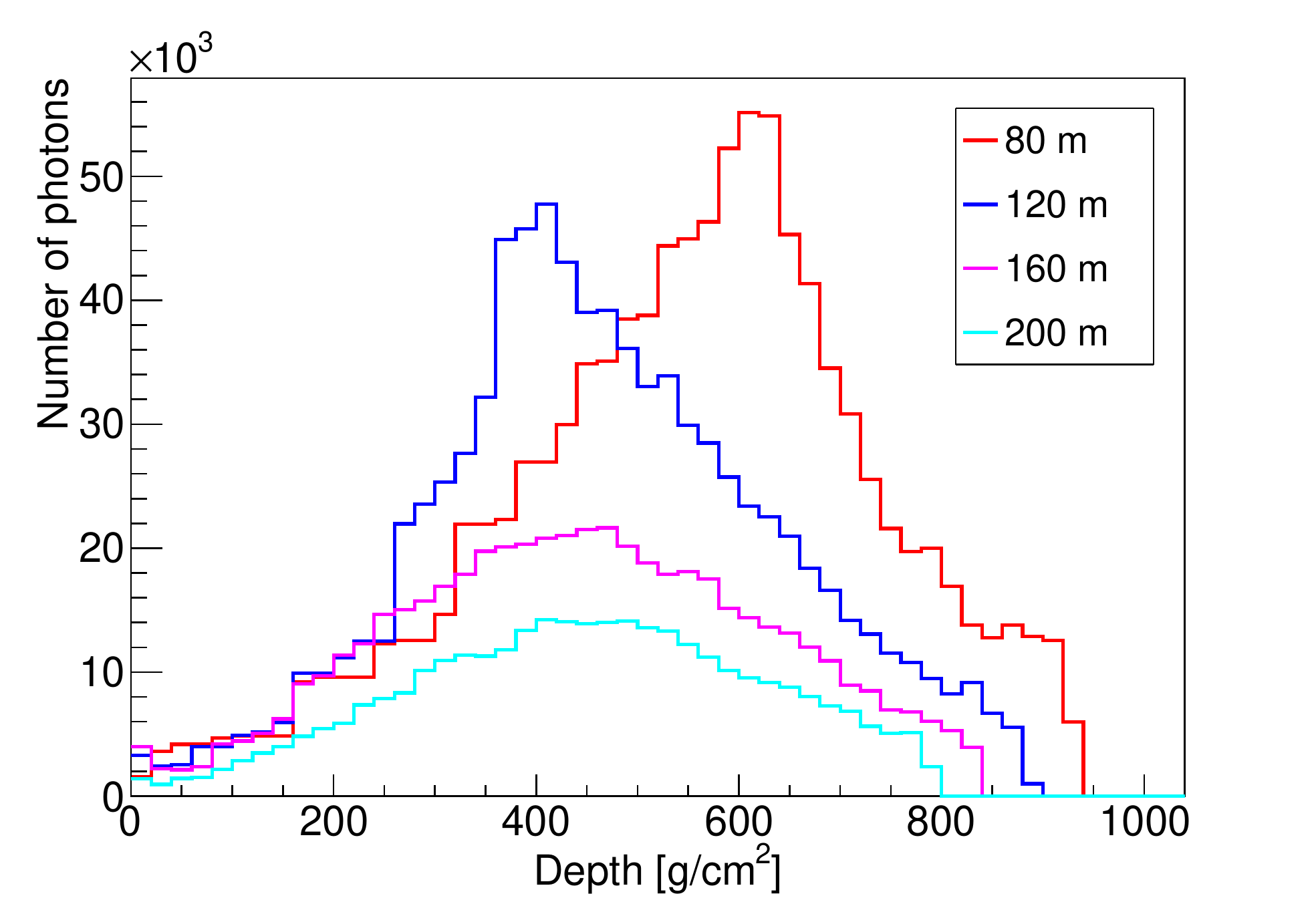}
  \caption{Cherenkov photon profiles of the same air shower event initiated by a proton of 100 TeV. Each curve represents the profile observed by telescopes at different distances from impact point.}
  \label{fig:long:Che:tels}
\end{figure}

Given that the Cherenkov emission is highly beamed, the Cherenkov photons emitted at a given depth can only be measured by telescopes in specific positions. In other words, each telescope samples different parts of the longitudinal development of the shower as shown in figure~\ref{fig:long:Che:tels} for telescopes at 80, 120, 160 and 200 m away from the shower core and as carefully discussed in reference~\cite{Giler:2021hvw}.

According to the model developed in reference~\cite{Arbeletche:2020rev}, the number of Cherenkov photons emitted in a given angle, $\theta$, with the shower axis by the particles in a shower can be written as
\begin{equation}
  \frac{\text{d}^2N_\gamma }{\text{d}\theta \; \text{d}X} (\theta,s,h)
   = \frac{1}{\pi} \, N_\text{e}(s) \times \sin \theta \times I(\theta,h) \times K(\theta,s,h)
\label{eq:app:final}
\end{equation}
where $s$ is age\footnote{$s = 3X/(X + 2\xmax)$}, $h$ height and $N_\text{e}$ the number of electrons in the shower. $I(\theta,h)$ and $K(\theta,s,h)$ are functions describing the angular emission of Cherenkov photons for which a parametrization is given in reference~\cite{Arbeletche:2020rev}. This parametrization describes quite well the behavior of the Cherenkov light distribution at short distances with respect to the shower core.

\begin{figure}[]
  \centering
  \includegraphics[scale=0.55]{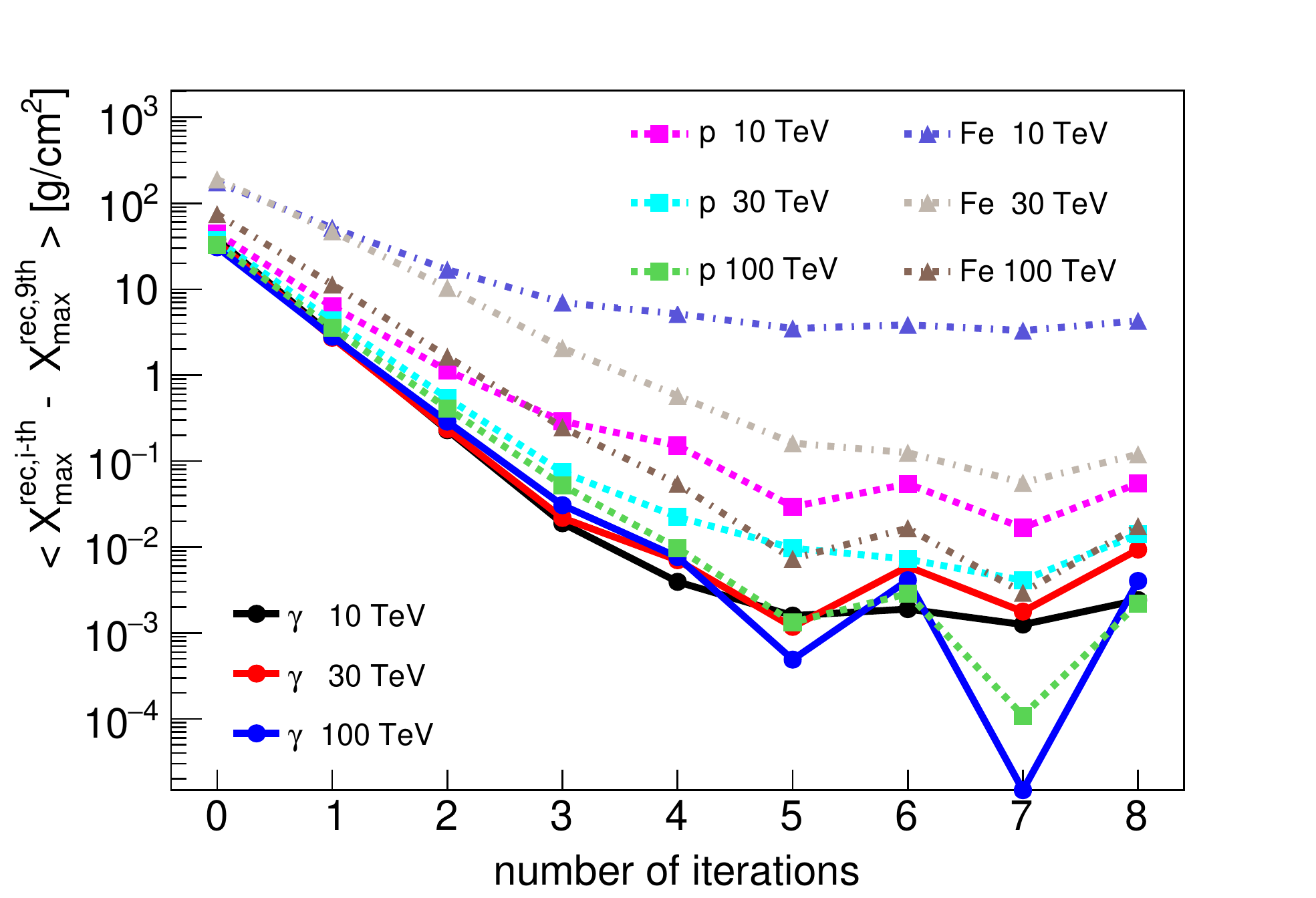}
  \caption{Mean value of \xmaxrec{} difference  at each step of iteration with respect to the \xmaxrec{} at 9th iteration.}
  \label{fig:convergence}
\end{figure}

Given the position of the telescope, equation~\ref{eq:app:final} allows us to calculate the number of electrons corresponding to the number of photons detected by each pixel. The calculation must be done in an iterative procedure given the dependence on age. We start the calculation considering $s=1$ for all pixels and find the corresponding longitudinal profile. We find the maximum of this longitudinal profile and recalculate the longitudinal profile using the new estimate of age. The procedure is repeated until convergence. Figure~\ref{fig:convergence} shows that after the 5th iteration the reconstructed depth in which the shower has the largest number of electrons, \xmaxrec, is stable within less than 1 \gcm.

\begin{figure}[]
  \centering
  \includegraphics[scale=0.52]{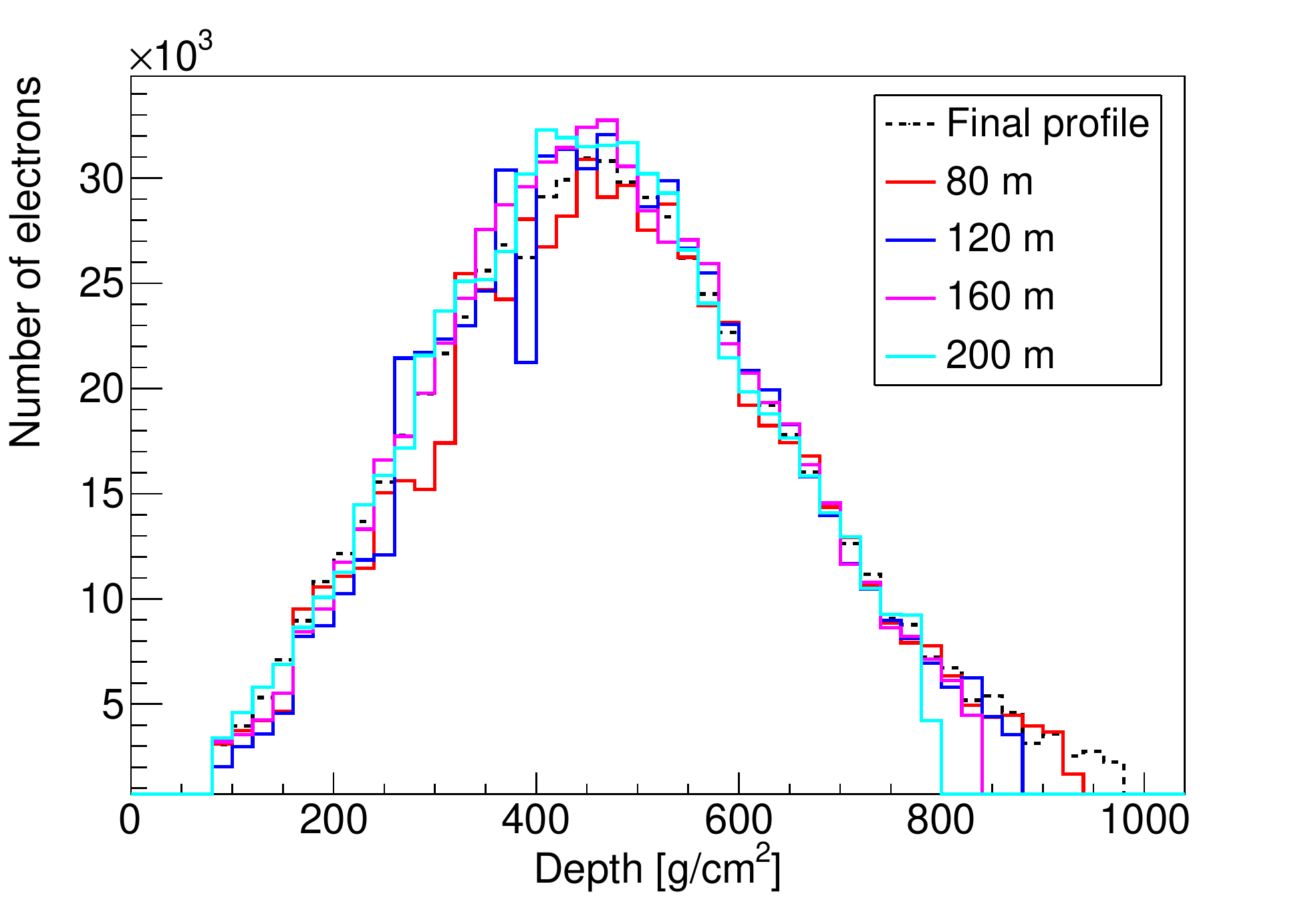}
  \caption{Electron shower profiles reconstructed at 9th iteration from Cherenkov photons profiles for telescopes at different distances. The final profile is the weighted average not only of the profiles shown, but of the complete telescope configuration.}
  \label{fig:long:rec}
\end{figure}

\begin{figure}[]
  \begin{center}
    \includegraphics[scale=0.55]{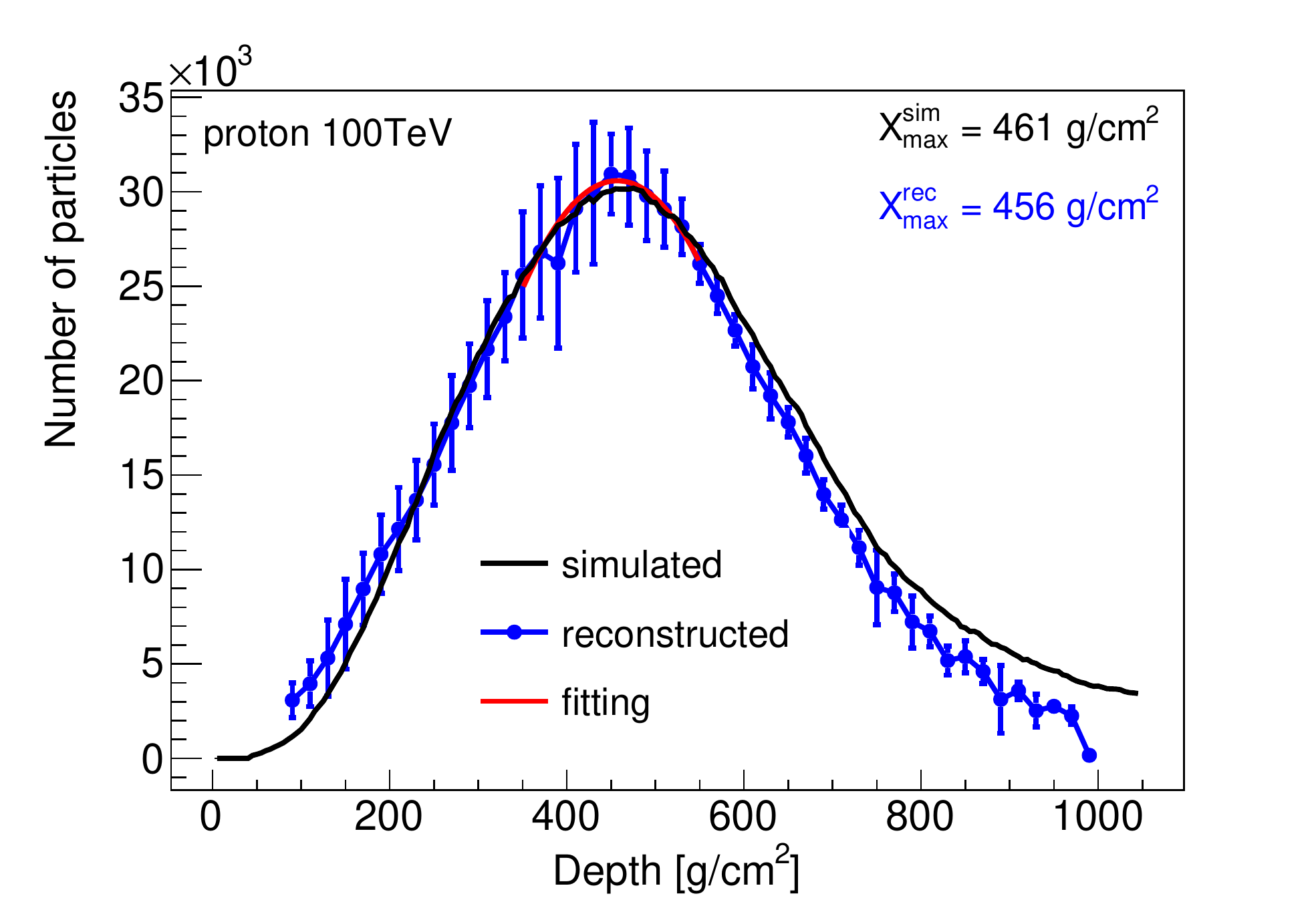}
  \end{center}
  \caption{Example of the simulated (black) and the corresponding reconstructed profile (blue). This event corresponds to an air shower initiated by a vertical proton of 100 TeV. The red curve is the second-degree polynomial fitting of the reconstructed profile. }
  \label{fig:xmax:rec}
\end{figure}

Figure~\ref{fig:long:rec} shows the longitudinal profile of electrons reconstructed using the procedure explained above. The number of photons simulated in each pixel of the 24 telescopes was projected into the shower plane and then projected into the shower axis. The longitudinal profile of Cherenkov photons was transformed into the number of electrons using equation~\ref{eq:app:final}. The contribution of the telescopes at 80, 120, 160 and 200 m away from the shower core are shown for comparison, however, the final reconstructed longitudinal electron profile includes the signal of all 24 telescopes in the simulation. \xmaxrec is calculated by fitting the reconstructed profile with a second-order polynomial ranging $\pm$100 \gcm around the peak as shown in figure~\ref{fig:xmax:rec}. The agreement between the reconstructed and simulated longitudinal electron profile is remarkable as well as the simulated (\xmaxsim) and reconstructed shower maximum.  Because the telescopes are pointing to zenith and have 10.5 degrees of field of view, they don’t detect Cherenkov photons generated deep in the atmosphere causing the reconstructed profile to have systematically less particles than the simulated profile at large depths.
\section{Results}
\label{sec:results}

\begin{figure}[]
  \centering
  \includegraphics[scale=0.65]{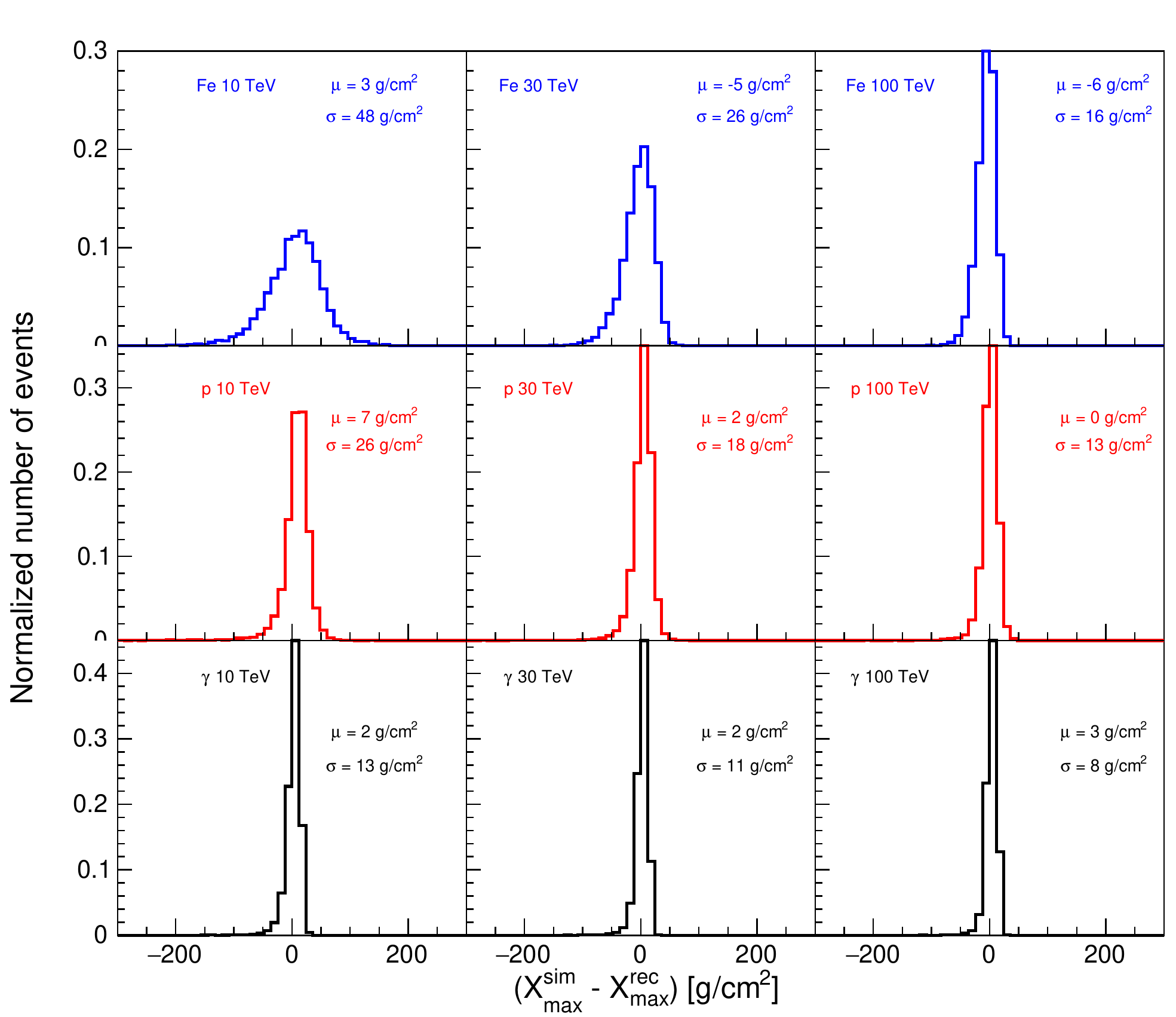}
  \caption{Distributions of $\xmaxsim - \xmaxrec$ for all primary particles and energies.}
  \label{fig:distributions}
\end{figure}

\begin{figure}[]
  \centering
  \includegraphics[scale=0.55]{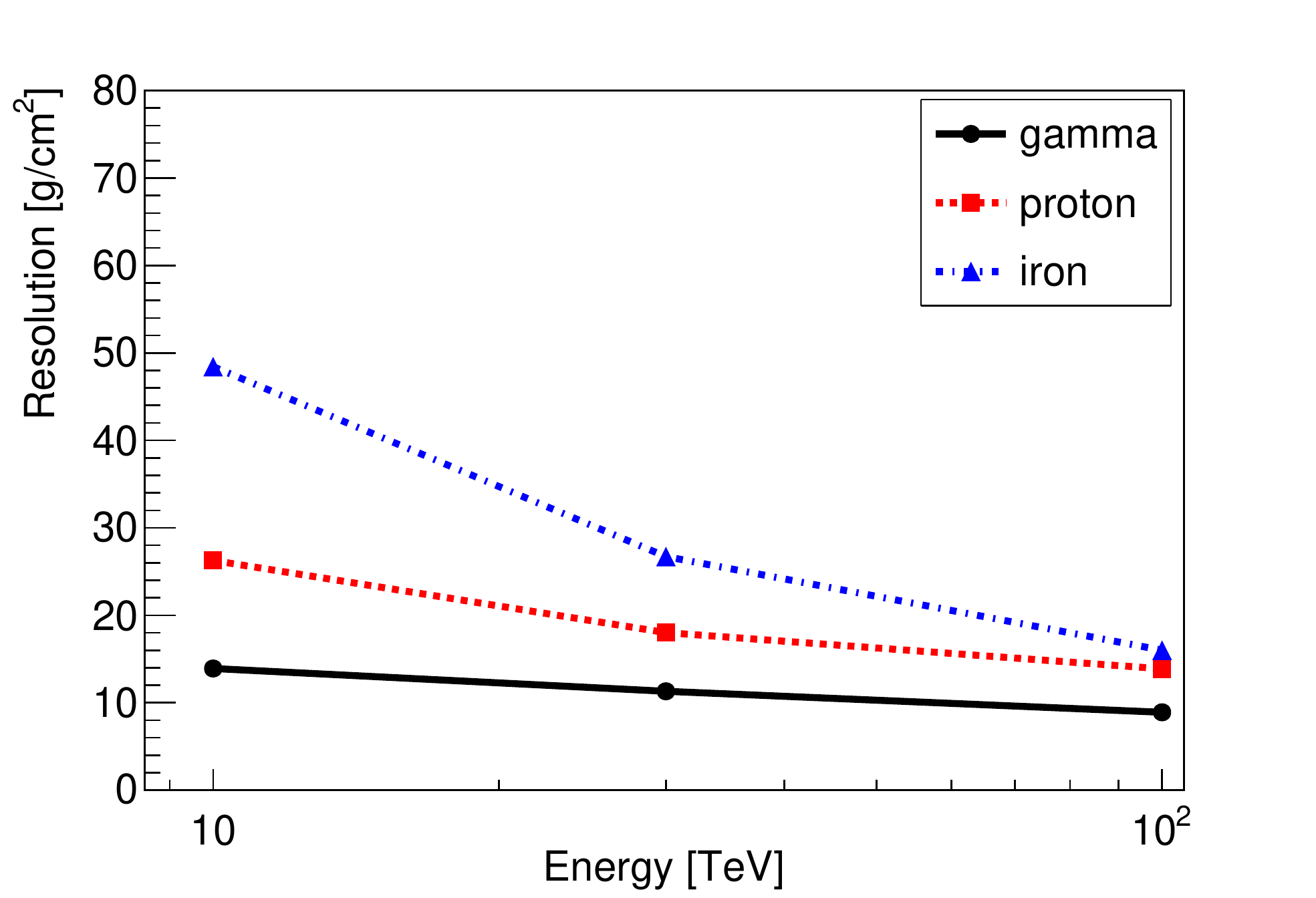}
  \caption{Resolution of the reconstruction, defined as the standard deviation of the distribution of  $\xmaxsim - \xmaxrec$ , as function of the primary energy for three different kinds of air showers.}
  \label{fig:resolution}
\end{figure}

Figure~\ref{fig:distributions} shows the distribution of $\xmaxsim - \xmaxrec$ for all primary particles and energies considered here. The distributions are narrow showing the quality of the method. The resolution of the reconstruction is taken as the standard deviation of these distributions and they are shown in figure~\ref{fig:resolution}.

For gamma-ray showers, the resolution is around 10 \gcm along the energy range. A deterioration of the resolution can be seen for cosmic ray events at low energies. In the best case at 100 TeV, the resolution for cosmic ray showers is below 16 \gcm.

The arrival direction accuracy (angular resolution) is probably the most important source of systematic uncertainty for this method. If the arrival direction is badly reconstructed, the projection of each pixel onto the shower plane would be systematically misplaced affecting the reconstruction of the longitudinal profile and therefore of \xmax. IACT angular resolution is typically 0.1 degrees and it is expected to be smaller than 0.05 degrees for CTA~\cite{science:with:cta}. We calculated the resolution in \xmax for 100 TeV proton events assuming that the reconstructed shower direction is 0.1 degrees apart from the simulated shower direction. The \xmax resolution worsen only by 1 \gcm.


\section{Conclusions}
\label{sec:conclusions}

In this paper, a novel method for the reconstruction of longitudinal air shower profiles using Cherenkov telescopes was presented. Using shower and simplified detector simulations, we have shown that the $\mathrm{X_{max}}$ parameter can be reliably estimated for each event. The resolution in the \xmax reconstruction, which was defined as the standard deviation of the difference distribution between \xmax simulated and \xmax reconstructed, was determined as a function of energy and type of primary particle. For all primaries, the resolution improves with energy. Moreover, the resolution is below 16 \gcm at 100 TeV for all studied cases. The resolution is worse for heavier nuclei which might indicate the need of a two-dimensional analysis for these primaries. The resolution achieved here makes it possible to study cosmic ray composition with Cherenkov telescopes.

Besides reconstructing \xmax, the method proposed here reconstruct the longitudinal development of electrons in the shower. This feature of the method is going to be developed in a future study towards reconstructing the energy of the shower. As it is widely used in the cosmic ray data analysis~\cite{bib:xmax:auger,bib:xmax:hires}, fitting the entire longintudinal development with a Gaisser-Hillas fuction~\cite{bib:gaisser:hillas} leads to a very accurate and bias free reconstruction of the shower energy.

The method developed here needs to be optimized for telescope type and the telescope array configuration. Exposure and acceptance estimations as well as systematic uncertainties should be considered when analysing real data towards composition studies. Moreover, the simplicity of the method proposed here, allow the use of the data from dozens of telescopes without the computational expenses required by template or deep learning techniques. It is expected that a fine-tuning of the method to the characteristics of each observatory will improve the resolution of \xmax predicted here. The \xmax resolution of gamma-rays, proton and iron nuclei at 100$\,$TeV are 8, 13, 16 \gcm, respectively. These resolutions are much smaller than the differences between the means of \xmax distributions, suggesting a real possibility to study cosmic rays with IACTs as complementary measurements together with other well-developed methods.

\section{Acknowledgments}
Authors acknowledge FAPESP Project 2015/15897-1 and the National Laboratory for Scientific Computing (LNCC/MCTI,  Brazil) for providing HPC resources of the SDumont supercomputer (http://sdumont.lncc.br). VdS acknowledges CNPq. This study was financed in part by the Coordenação de Aperfeiçoamento de Pessoal de Nível Superior - Brasil (CAPES) - Finance Code 001. The authors thank Luan B. Arbeletche for fruitful discussions.

\bibliography{references}

\newpage

\end{document}